# Quasi-analytical solutions of hybrid platform and the optimization of highly sensitive thin-film sensors for terahertz radiation

**PIYAWATH TAPSANIT,[1] MASATSUGU YAMASHITA,[1] TERUYA ISHIHARA,[2]**

**CHIKO OTANI,[1,2,*]**

[1]*RIKEN Center for Advanced Photonics, Sendai, 980-0845, Japan*
[2]*Department of Physics, Tohoku University, Sendai, 980-8578, Japan*
*\*Corresponding author: otani@riken.jp*

We present quasi-analytical solutions (QANS) of hybrid platform (HP) comprising metallic grating (MG) and stacked-dielectric layers for terahertz (THz) radiation. The QANS are validated by finite difference time domain simulation. It is found that the Wood's anomalies induce the high-order spoof surface plasmon resonances in the HP. The QANS are applied to optimize new perfect absorber for THz sensing of large-area thin film with ultrahigh figure of merit reaching fifth order of magnitude for the film thickness 0.0001p (p:MG period). The first-order Wood's anomaly of the insulator layer and the Fabry-Perot in the slit's cavity account for the resonance of the perfect absorber. The QANS and the new perfect absorber may lead to highly sensitive and practical nano-film refractive index sensor for THz radiation.

http://dx.doi.org/10.1364/AO.99.09999

## 1. Introduction

The terahertz (THz) radiation lies between the microwave and the infrared radiation. It is applied broadly to the areas spanning from food quality control, medical care and security sectors to fundamental science and cultural heritage [1]. Especially, the THz light promises non-invasive sensing and imaging of biomedical objects due to its non-ionizing energy, 1 THz = 4.14 meV [2]. In sensing, the sensitivity of a sensor to the variations of sample's properties can be enhanced by using plasmonic platforms such as metallic grating (MG) [3] or metallic hole arrays (HA) [4,5]. The enhanced sensitivity of these structures arise due to the coupled surface plasmon polaritons (SPPs) resonances generated on two faces of the apertures in optical band [6,7] and the coupled spoof surface plasmon polaritons (SSPPs) resonances whose properties depend on apertures' geometry in the THz band [8]. The HA sensors' sensitivity is more enhanced by isolating plasmonic resonances of the HA from those of substrate using the hybrid substrate (HS) [9] or with the transverse magneto-optical Kerr effect in hybrid structure of the HA and the Cobalt layer [10]. The responses of the HA sensors to the variations of properties of bulk medium can be analytically investigated by the coupled-mode analysis [11] or scattering-matrix approach [10]. However, the analytical solutions of these hybrid structures with the presence of thin films are still lacking.

Furthermore, the subwavelength focusing can be achieved by defining the HS as alternating metal/insulator layers forming hyperbolic metamaterial [12]. However, the conventional theory always assumes the apertures as the point sources neglecting the effects of apertures' geometry on focusing features. Therefore, we are forced to employ the numerical methods such as finite difference time domain (FDTD) method [13] and commercial finite element software (e.g., CST STUDIO and COMSOL) to optimize these structures. However, these methods are time-consuming and inaccurate when the structures are much smaller than the light wavelength hindering the optimizations of THz subwavelength sensing and focusing using HA and MG.

Here, we propose the quasi-analytical solutions (QANS) of the hybrid platform (HP) comprising the MG and stacked-dielectric layers (SDLs) attached on both sides of the MG to provide a deeper understanding of HP's resonance features and the efficient tools for the optimization of HP structures for the applications in subwavelength sensing and focusing with THz radiation. As an example of the application of QANS, we propose and optimize a new perfect absorber for detecting the variations of refractive index and thickness of a thin-film by measuring the change of reflectance of the perfect absorber. The new perfect absorber shows ultrahigh figure of merit which is dependent on the refractive index and thickness of a sample layer. The QANS allow us to mathematically understand the physical mechanism of this perfect absorber and they may also lead to more practical thin-film refractive index sensor for THz radiation.

## 2. Method: QANS of hybrid platform

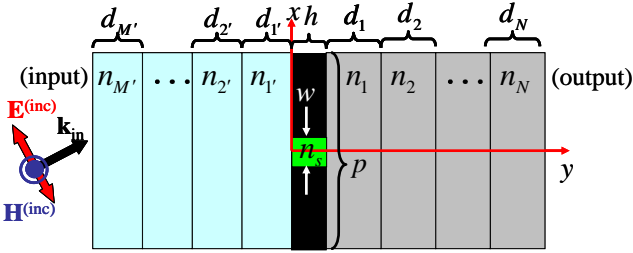

Fig. 1. Schematic view of the HP comprising the MG labelled as black (metallic part) and green (slit part) regions and SDLs attached on two faces of the MG. The relevant parameters are indicated in the figure.

The schematic view of the HP is drawn in Fig. 1 where the metallic part of the MG is labelled by the black color, the isotropic material with refractive index $n_s$ and dielectric constant $\varepsilon_s$ filling the slit is labelled by the green color. The isotropic dielectric layers with refractive indices indicated in Fig. 1 (dielectric constants have the same subscripts as those of the refractive indices) comprising the SDLs are attached on two faces of the MG. The SDLs on the input side may be considered as the hybrid substrate and the SDLs on the output side may be considered as the stacked films and vice versa. The primes are used to distinguish the SDLs in the input side from those in the output side. The transverse magnetic (TM) wave with the amplitude of the incident magnetic field $H_0$ is incident on the HP from the input region. The metal is modeled as the perfect electric conductor which is valid for THz and lower frequency ranges because the light wavelength is much longer than the metal skin depth (about 61 nm for gold at 1 THz [14]) so that light does not penetrate into the metal. The slit filled with isotropic material with refractive index $n_s$ has $w\ll\lambda$ so that only the slit fundamental waveguide mode denoted in real-space as $\langle x|0\rangle = 1/\sqrt{w}$ is excited within the slit's cavity [11]. The solutions of $H_z$ in remaining regions are expanded by Bloch basis functions denoted as $\langle x|\beta_m\rangle = \exp(i\beta_m x)/\sqrt{p}$, where $\beta_m = k_x + m2\pi/p$, $k_x$ is tangential component of the incident wavevector, $m$ is an integer, and $p$ is period of the MG with field coefficients $A_m^{(\alpha)}$ and $B_m^{(\alpha)}$ for $+y$ and $-y$ travelling waves, respectively, in region $\alpha$. The electric field is obtained by $H_z$ using the Maxwell and Amperes' law. The continuities of $H_z$ and $E_x$ at the interfaces between two regions lead to the $m$th-order reflection coefficient denoted as $B_m^{(\text{in})}/H_0$ and the $m$th-order transmission coefficient denoted as $A_m^{(\text{out})}/H_0$ written in Eq. (1) and (2), respectively, where $\phi_0^{(\text{in})} = q_0^{(\text{in})} D_{\text{in}}$, $D_{\text{in}}$ is the total thickness of SDLs in the input side, $\delta_{m0}$ is Dirac delta function, $I_0 = -\langle \beta_0|0\rangle$ denotes direct excitation, $Y_m^{(\alpha)} = k_\alpha/q_m^{(\alpha)}$ is admittance, $q_m^{(\alpha)} = \sqrt{k_\alpha^2 - \beta_m^2}$ is normal component of Bloch wavevectors, $\gamma = \cot(k_s h)$ accounts for the multiple reflections in the slit's cavity, $G_V = \csc(k_s h)$ accounts for the coupling of the diffraction modes on two faces of the MG through the slit, $\langle \beta_m|0\rangle = \sqrt{w/p}\,\text{sinc}(\beta_m w/2)$ is direct product between the Bloch basis function and fundamental waveguide mode, and $k_\alpha$ is wavenumber in region $\alpha$.

$$\frac{B_m^{(\text{in})}}{H_0} = \exp(-i\phi_0^{(\text{in})})\left(\frac{\overline{S}_{22}^{(0)}+\overline{S}_{21}^{(0)}}{\overline{S}_{12}^{(0)}+\overline{S}_{11}^{(0)}}\right)\left\{\delta_{m0} + \left(\frac{t_{\text{in},M'}^{(0)}\cdots t_{2',1'}^{(0)}}{\overline{S}_{22}^{(0)}+\overline{S}_{21}^{(0)}}\right)I_0 \frac{n_s\varepsilon_{1'}}{n_{1'}\varepsilon_s}\right.$$
$$\left.\times iY_m^{(1')}\frac{2\det(\overline{\mathbf{S}}^{(m)})}{t_{\text{in},M'}^{(m)}\cdots t_{2',1'}^{(m)}(\overline{S}_{12}^{(m)}+\overline{S}_{11}^{(m)})}\langle\beta_m|0\rangle\frac{G_{\infty S}^{(\text{out})}-\gamma}{\Omega}\right\},\quad(1)$$

$$\frac{A_m^{(\text{out})}}{H_0} = \exp(-i\phi_0^{(\text{in})})\left(\frac{t_{\text{in},M'}^{(0)}\cdots t_{2',1'}^{(0)}}{\overline{S}_{12}^{(0)}+\overline{S}_{11}^{(0)}}\right)I_0 \frac{n_s\varepsilon_1}{n_1\varepsilon_s}$$
$$\times iY_m^{(1)}\frac{2\det(\mathbf{S}^{(m)})}{t_{N,N-1}^{(m)}\cdots t_{2,1}^{(m)}(S_{22}^{(m)}+S_{21}^{(m)})}\langle\beta_m|0\rangle\frac{G_V}{\Omega}.\quad(2)$$

The $\Omega$ is defined as the dispersion function written as follows
$$\Omega = \left(G_{\infty S}^{(\text{in})}-\gamma\right)\left(G_{\infty S}^{(\text{out})}-\gamma\right)-G_V^2,\quad(3)$$

where $G_{\infty S}^{(\text{in})}$ and $G_{\infty S}^{(\text{out})}$ describe the re-radiation and re-absorption of diffraction modes by the fundamental waveguide modes [11] defined here as follows

$$G_{\infty S}^{(\text{in})} = i\frac{n_s\varepsilon_{1'}}{n_{1'}\varepsilon_s}\sum_{m=-\infty}^{\infty}\left(\frac{\overline{S}_{11}^{(m)}-\overline{S}_{12}^{(m)}}{\overline{S}_{11}^{(m)}+\overline{S}_{12}^{(m)}}\right)Y_m^{(1')}\langle 0|\beta_m\rangle\langle\beta_m|0\rangle,\quad(4)$$

$$G_{\infty S}^{(\text{out})} = i\frac{n_s\varepsilon_1}{n_1\varepsilon_s}\sum_{m=-\infty}^{\infty}\left(\frac{S_{22}^{(m)}-S_{21}^{(m)}}{S_{22}^{(m)}+S_{21}^{(m)}}\right)Y_m^{(1)}\langle 0|\beta_m\rangle\langle\beta_m|0\rangle,\quad(5)$$

The $G_{\infty S}^{(\text{in})}$ and $G_{\infty S}^{(\text{out})}$ are rapidly converged with the convergent error defined as 0.001% throughout the paper. The dispersion relation is obtained by the condition $\text{Re}(\Omega)=0$. The scattering matrices are defined as follows
$$\overline{\mathbf{S}}^{(m)} = \overline{\mathbf{T}}_{\text{in},M'}^{(m)}\cdots\overline{\mathbf{T}}_{2',1'}^{(m)},\ \mathbf{S}^{(m)} = \mathbf{T}_{\text{out}}^{(m)}\mathbf{T}_{N,N-1}^{(m)}\cdots\mathbf{T}_{2,1}^{(m)},\quad(6)$$

where $\overline{\mathbf{T}}_{\alpha',\beta'}^{(m)}$, $\mathbf{T}_{\alpha,\beta}^{(m)}$, and $\mathbf{T}_{\text{out}}^{(m)}$ are transfer matrices defined in Eq. (7), where $\phi_m^{(\alpha)} = q_m^{(\alpha)}d_\alpha$, $r_{\alpha,\beta}^{(m)}$ and $t_{\alpha,\beta}^{(m)}$ are reflection and transmission coefficients of TM waves given in Eq. (8).

$$\overline{\mathbf{T}}_{\alpha',\beta'}^{(m)} = \begin{bmatrix} e^{-i\phi_m^{(\beta')}} & r_{\alpha',\beta'}^{(m)}e^{i\phi_m^{(\beta')}} \\ r_{\alpha',\beta'}^{(m)}e^{-i\phi_m^{(\beta')}} & e^{i\phi_m^{(\beta')}} \end{bmatrix},\ \mathbf{T}_{\alpha,\beta}^{(m)} = \begin{bmatrix} e^{i\phi_m^{(\beta)}} & r_{\alpha,\beta}^{(m)}e^{-i\phi_m^{(\beta)}} \\ r_{\alpha,\beta}^{(m)}e^{i\phi_m^{(\beta)}} & e^{-i\phi_m^{(\beta)}} \end{bmatrix}$$
$$,\ \mathbf{T}_{\text{out}}^{(m)} = \begin{bmatrix} t_{N,\text{out}}^{(m)}e^{i\phi_m^{(N)}} & 0 \\ -r_{N,\text{out}}^{(m)}e^{i\phi_m^{(N)}} & e^{-i\phi_m^{(N)}} \end{bmatrix}.\quad(7)$$

$$r_{\alpha,\beta}^{(m)} = \frac{\varepsilon_\beta q_m^{(\alpha)}-\varepsilon_\alpha q_m^{(\beta)}}{\varepsilon_\beta q_m^{(\alpha)}+\varepsilon_\alpha q_m^{(\beta)}},\ t_{\alpha,\beta}^{(m)} = \frac{2\varepsilon_\beta q_m^{(\alpha)}}{\varepsilon_\beta q_m^{(\alpha)}+\varepsilon_\alpha q_m^{(\beta)}}.\quad(8)$$

In the output side, the field coefficients are obtained by $\left( A_m^{(\text{out})} \ 0 \right) = \mathbf{T}_{\text{out}}^{(m)} \left( A_m^{(N)} \ B_m^{(N)} \right)$ and $\left( A_m^{(\alpha+1)} \ B_m^{(\alpha+1)} \right) = \mathbf{T}_{\alpha+1,\alpha}^{(m)} \times \left( A_m^{(\alpha)} \ B_m^{(\alpha)} \right) / t_{\alpha+1,\alpha}^{(m)}$. In the input side, the field coefficients are obtained by $\left( H_0 \delta_{m0} \exp\left(-i\phi_0^{(\text{in})}\right) \ B_m^{(\text{in})} \right) = \overline{\mathbf{T}}_{\text{in},M'}^{(m)} \times \left( A_m^{(M')} \ B_m^{(M')} \right) / t_{\text{in},M'}^{(m)}$, and $\left( A_m^{(\alpha'+1)} \ B_m^{(\alpha'+1)} \right) = \overline{\mathbf{T}}_{\alpha'+1,\alpha'}^{(m)} \times \left( A_m^{(\alpha')} \ B_m^{(\alpha')} \right) / t_{\alpha'+1,\alpha'}^{(m)}$. Note that the matrix transpose is implied in the matrix multiplication. Finally, the reflectance $R$ and transmittance $T$ are computed by Eq. (9), where the function P includes only the propagating waves.

$$R = \sum_m P\left\{ \frac{q_m^{(\text{in})}}{q_0^{(\text{in})}} \left| \frac{B_m^{(\text{in})}}{H_0} \right|^2 \right\}, \ T = \frac{\varepsilon_{\text{in}}}{\varepsilon_{\text{out}}} \sum_m P\left\{ \frac{q_m^{(\text{out})}}{q_0^{(\text{in})}} \left| \frac{A_m^{(\text{out})}}{H_0} \right|^2 \right\}. \quad (9)$$

## 3. Comparison between QANS and FDTD of air/dielectric/MG/dielectric/air hybrid platform

In order to validate the QANS, we show the comparison between QANS and FDTD transmission spectra of a simple air/HR-Si/MG/Teflon/air HP ($M=N=1$) for $k_x=0$ in Fig. 2. The imaginary parts of the dielectric constants of high resistivity Si (HR-Si) and Teflon are neglected to reduce the simulation time of the FDTD and the absorption will be given by using only QANS. It can be seen that the FDTD spectra approach the QANS spectrum by increasing the resolutions from 3 to 10 pix/μm indicating that the QANS are consistent with FDTD with high resolution. The QANS resonances are at 2.1325 THz, 2.7676 THz, and 2.8860 THz which can be precisely obtained by Re(Ω)=0. Then, the transmittance at the resonant peak is directly proportional to $1/|\text{Im}(\Omega)|^2$ according to Eq. (2) and (9). The good way to understand these resonances is by first looking at the Wood's anomalies [17] at which $1/|\Omega|\approx 0$. The Wood's anomalies generate the short circuit resulting in the zero transmittance at the frequencies 2.0495 THz, 2.6872 THz, and 2.8546 THz. The Wood's anomalies arise when $\left| \text{Re}\left(G_{\infty s}^{(\text{in})}\right) \right| \to \infty$ or $\left| \text{Re}\left(G_{\infty s}^{(\text{out})}\right) \right| \to \infty$ with finite $\gamma$ and $G_V$. The latter explodes due to the interference between 0th-order and ±$m$th-orders diffraction modes resulting in the following divergent condition

$$q_0^{(1)} q_{\pm m}^{(1)} \left[ 1 - r_{1,\text{out}}^{(0)} \exp\left(2i\phi_0^{(1)}\right) \right] \left[ 1 - r_{1,\text{out}}^{(\pm m)} \exp\left(2i\phi_{\pm m}^{(0)}\right) \right] = 0. \quad (10)$$

The Eq. (10) is satisfied if $q_n^{(1)} = 0$, where $n=0,\pm m$, corresponding to the Rayleigh's anomalies in which the diffraction waves propagate exactly parallel to the tangential axis [18]. This equation is also satisfied if $r_{1,\text{out}}^{(n)} \exp\left(2i\phi_n^{(1)}\right) = 1$ corresponding to the $n$th-order Wood's anomaly and can be fulfilled when the $n$th-order diffraction mode is propagating wave in the dielectric layer ($|k_1| > |\beta_n|$) but becomes evanescent wave in the output medium ($|k_{\text{out}}| < |\beta_n|$). By imposing these requirements, we finally obtain the simplified condition of the $n$th-order Wood's

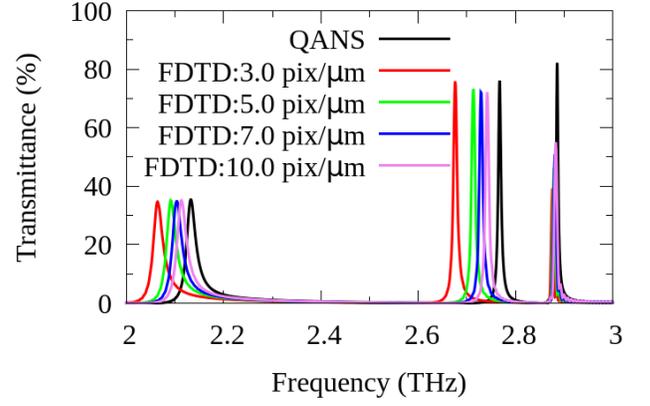

Fig. 2. QANS and FDTD transmission spectra of air/HR-Si/MG/Teflon/air HP ($M=N=1$). The slit is filled with air. The dielectric constants of HR-Si and Teflon are 11.7 [15] and 2.088 [16] with neglected losses, respectively. $d_1=d_{1'}=10$ μm, $w=h=1$ μm, and $p=100$ μm. The structure is excited by the normal incidence light ($k_x=0$).

anomaly as follows

$$\tan\left(d_1 \sqrt{k_1^2 - \beta_n^2}\right) = \left(\varepsilon_1 / \varepsilon_{\text{out}}\right) \sqrt{\left(\beta_n^2 - k_{\text{out}}^2\right) / \left(k_1^2 - \beta_n^2\right)}. \quad (11)$$

The Wood's anomaly from the divergence of $G_{\infty s}^{(\text{in})}$ has the same form as Eq. (11) with the substitutions of dielectric constants, dielectric layer thickness, and wavenumbers: $\varepsilon_1 \to \varepsilon_{1'}$, $\varepsilon_{\text{out}} \to \varepsilon_{\text{in}}$, $d_1 \to d_{1'}$, $k_1 \to k_{1'}$, and $k_{\text{out}} \to k_{\text{in}}$. Therefore, the Wood's anomalies at 2.0495 THz and 2.6872 THz are the 1st-order and 2nd-order in HR-Si layer, respectively, while the Wood's anomaly at 2.8546 THz is the 1st-order in Teflon layer as indicated by labels in Fig. 3. The Wood's anomalies trigger the phase oscillation of the diffraction term of the reflection coefficient, that is the second term in the curly bracket of Eq. (1), which becomes out of phase to the non-diffraction term, that is the first term in the curly bracket of Eq. (1) resulting in the transmission peaks at the resonances. The resonant positions correspond to the crossing points between the two components of Ω which occur just next to the Wood's anomalies as indicated by the circles in Fig. 3. Therefore, the resonances at 2.1325 THz, 2.7676 THz and 2.8860 THz are induced by the 1st-order Wood's anomaly in HR-Si, the 2nd-order Wood's anomaly in HR-Si, and the 1st-order Wood's anomaly in Teflon, respectively. In other words, these resonances cannot arise if their associated $m$th-order diffraction modes which cause the $m$th-order Wood's anomalies are missing from the coupling parameters $G_{\infty s}^{(\text{in})}$ (for HR-Si) and $G_{\infty s}^{(\text{out})}$ (for Teflon). For example, the resonances at 2.1325 THz and 2.7676 THz are demolished following the disappearances of the 1st-order Wood's anomaly and the 2nd-order Wood's anomaly, respectively, in HR-Si by discarding the $m=\pm 1$ and $m=\pm 2$ diffraction modes, respectively, in $G_{\infty s}^{(\text{in})}$ as shown by the green and cyan dashed lines, respectively, in Fig. 3. This argument is also valid for $k_x \neq 0$. Indeed, the dispersion relation of the resonances retrieved by the condition Re(Ω)=0 follow the dispersion relation of the Wood's anomalies obtained by Eq. (11) (not shown).

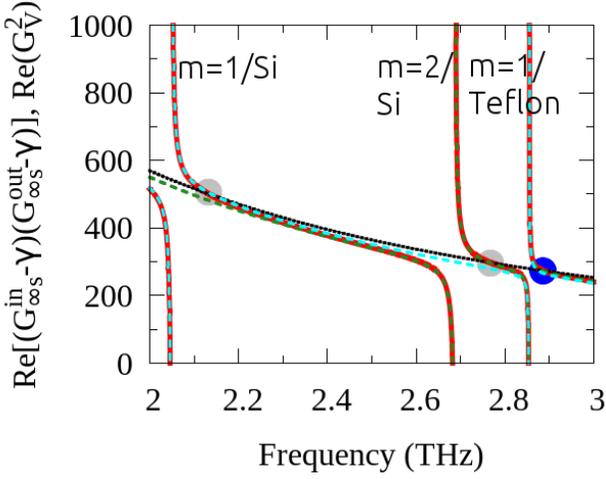

Fig. 3. Real parts of the components of $\Omega$. The red solid line and the black dotted line denote $\mathrm{Re}\left(\left(G_{\infty s}^{(\mathrm{in})}-\gamma\right)\left(G_{\infty s}^{(\mathrm{out})}-\gamma\right)\right)$ and $G_V^2$, respectively, taking into account all diffraction modes. The grey and blue circle indicate the resonant positions induced by Wood's anomalies in HR-Si and Teflon, respectively. The green and cyan dashed lines denote $\mathrm{Re}\left(\left(G_{\infty s}^{(\mathrm{in})}-\gamma\right)\left(G_{\infty s}^{(\mathrm{out})}-\gamma\right)\right)$ with discarding $m=\pm1$ and $m=\pm2$ diffraction modes in $G_{\infty s}^{(\mathrm{in})}$, respectively.

We will now discuss some differences of QANS and FDTD spectra spotted in Fig. 2. The relatively large differences between the QANS and FDTD resonant frequencies of the first and second resonances from the lower-energy-side are due to the low resolution of the FDTD in high-dielectric-constant HR-Si layer. The resonant frequencies of the third resonant peaks from two methods are almost coincided due to the smaller dielectric constant of Teflon. However, the QANS transmittance at the third resonance is significantly larger than the FDTD transmittance. This behavior comes from the localized surface plasmons (LSPs) at the slit corners. It has been pointed out that the quenched transmittance is associated with the presence of the LSPs [19]. We will see this behavior clearer by considering the electric field distributions at all resonant frequencies shown in Fig. 4(a)-(c). The QANS $|\mathbf{E}|$ distributions at 2.1325 THz and 2.7676 THz are normalized by the maxima of $|\mathbf{E}|/E_0$, where $E_0$ is the amplitude of incident electric field which occur at $(x,y)=(0,0)$ inside the HR-Si layer and equal to 63.0 and 85.0, respectively, while the QANS $|\mathbf{E}|$ distribution at 2.8860 THz is normalized by the maximum of $|\mathbf{E}|/E_0$ which occurs at $(x,y)=(h,0)$ inside the Teflon layer and equal to 88.5. The larger electric fields at the slit openings are associated with the higher transmittance. The $|\mathbf{E}|$ distributions from QANS are consistent with those from FDTD except the contrast due to the excitations of LSPs to be described shortly. In Fig. 4(a), we see that the electric field from QANS and FDTD is localized on the surface of the HR-Si layer. The maximum of $\mathbf{E}|/E_0$ taken from QANS is 17.7 at $x=\pm28.10$ μm with the electric field decay length 21.53 μm. The localization with strong electric field enhancement results dominantly from the decay of the $m=\pm1$ diffraction modes in the input region. Therefore, we call the resonance at 2.1325 THz as the 1st-order SSPP on the surface of HR-Si layer with the shorthand SSPP(HR-Si,$m=1$). Likewise, we also see the localization of the electric field on the surface of the HR-Si layer in Fig. 4(b) with higher maximum of $|\mathbf{E}|/E_0$ 22.0 at $x=\pm14.15$ μm and the shorter electric field decay length 10.95 μm due mainly to the decay of $m=\pm2$ diffraction modes in the input region. Therefore, we name the resonance at 2.7676 THz as SSPP(HR-Si,$m=2$). The $|\mathbf{E}|$ distribution at 2.8860 THz becomes strongly localized on the surface of Teflon layer as shown in Fig. 4(c) with the maximum of $|\mathbf{E}|/E_0$ 36.8 at $x=\pm25.62$ μm and the electric field decay length 58.27 μm due to the decay of the $m=\pm1$ diffraction modes in the output region. Therefore, we call the resonance at 2.8860 THz as SSPP(Teflon,$m=1$). The electric field decay length of SSPP(Teflon,$m=1$) is longer than that of SSPP(HR-Si,$m=1$) because the dielectric constant of Teflon is smaller than that of HR-Si. We zoom in the $|\mathbf{E}|$ distributions around the slit regions as indicated by the white arrows in Fig. 4(a)-(c), and we find that the $|\mathbf{E}|$ distributions from FDTD are strongly localized at the slit corners, while the |E| distributions from QANS lack these features and the electric fields are maximum at the center of the slit's openings. These effects are known as LSPs found in metallic nanoparticles [20]. Therefore, the low contrast of QANS |E| distribution is associated with the lack of LSPs. The LSPs may be accounted for by including higher waveguide modes which are evanescent waves in the slit's cavity within this frequency range [21]. The higher waveguide modes would give larger Im($\Omega$) and thus the resonant transmission would decrease.

Next, we explain the effects of losses in dielectric layers. The QANS absorbance $A$ is determined by $A=1-R-T$. The red solid line in Fig. 4(d) shows the absorption generated by losses in both HR-Si and Teflon layers, and the absorbance at SSPP(HR-Si,$m=1$), SSPP(HR-Si,$m=2$), and SSPP(Teflon,$m=1$) are 4.8%, 40.4%, and 70.5%, respectively. The HP is heated mostly in the Teflon layer because of the small loss of HR-Si. By including only loss in Teflon, the absorbance at SSPP(Teflon,$m=1$) remains the same, while the absorbance at SSPP(HR-Si,$m=1$) and SSPP(HR-Si,$m=2$) decrease by 0.8% and 1.6%, respectively, from the absorbance of the former case. Notice that the high absorbance can be generated at SSPP(HR-Si,$m=2$) even without loss in HR-Si layer implying the coupling between SSPP(HR-Si, $m=2$) and SSPP (Teflon,$m=1$). This coupling manifests in the |E| distribution at the SSPP(HR-Si,$m=2$) in which the weaker excitation of SSPP(Teflon,$m=1$) can be also seen behind the grating. By including the higher waveguide modes to account for the LSPs, the absorbance is expected to increase following the quenched transmittance [19] because the strong electric fields around the slit corners in Teflon layer could additionally generate more heat.

Before leaving this section, we discuss the limitations of QANS. QANS assume the narrow slit width to take only the fundamental waveguide mode into account missing the LSPs which are excited at the slit corners. However, the QANS transmittance at at 2.1325 THz, corresponding to the wavelength about 141 μm, is almost identical to that from the FDTD which means that the QANS give accurate results

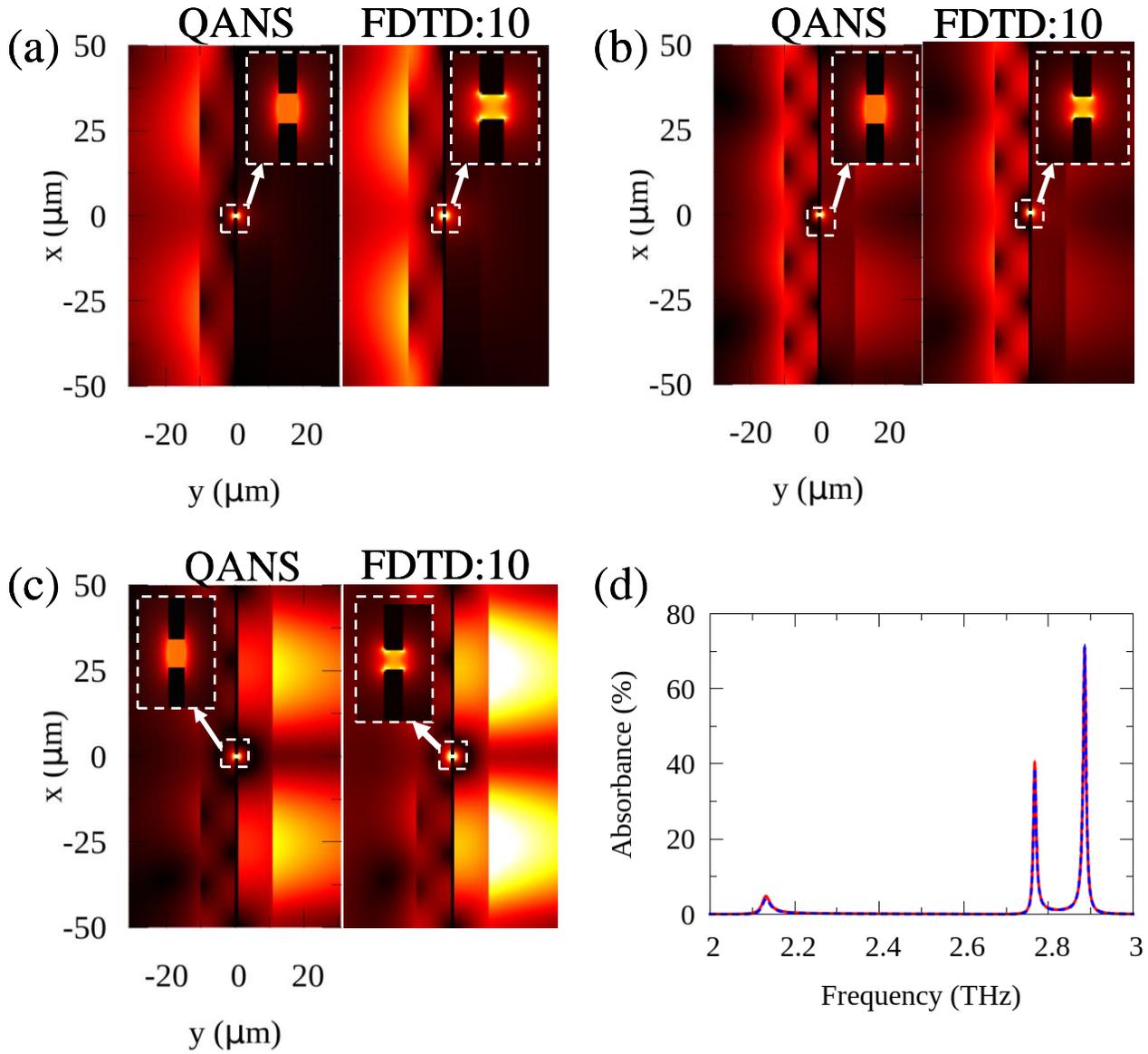

Fig. 4. (a) |**E**| distributions at 2.1325 THz for QANS and 2.1130 THz for FDTD: 10 pix/μm. (b) |**E**| distributions at 2.7676 THz for QANS and 2.7420 THz for FDTD: 10 pix/μm. (c) |**E**| distributions at 2.8860 THz for QANS and 2.8830 THz for FDTD: 10 pix/μm. (d) Absorption spectra due to the presence of losses in both HR-Si and Teflon layers (red solid line), and only in Teflon (blue dashed line). The imaginary part of the dielectric constants of HR-Si and Teflon are 0.0004 [15] and 0.02 [16], respectively.

When $w/\lambda \leq 7.09 \times 10^{-3}$. The transmission below this resonance behaves as the capacitor-like similar to that of the structure with slit width smaller than the skin-depth of the metal [22]. However, the upper limit of the slit width must be addressed rigorously, and we leave this task to the future work. There are two applications of QANS that can be immediately envisaged. The first application is the optimization of perfect absorbers where one of dielectric layers in the output side is defined as the metallic plane and lossy dielectric layers are sandwiched between the grating and the metallic plane, which will be discussed in the following section [23]. The second application is the optimization of the subwavelength focusing devices where the stacked-dielectric layers are defined as alternating metal/insulator layers forming the hyperbolic metamaterials to focus large-wavevector waves generated by the MG into the subwavelength volume, the technique useful for the optical lithography and subwavelength imaging [12]. The latter application will be discussed in the separated paper.

# 4. Applications : Optimization of highly sensitive large-area thin-film sensor for THz radiation

## A. QANS of input/I/MG/M-HP without and with a sample

The HP can be transformed into the widely-known structure of the perfect absorber made by sandwiching lossy insulator layer between the MG and thick metallic layer [23]. The equivalent HP structure of this absorber is input/MG/I/M/output HP ("I" for an insulator and "M" for a metal). The thick metallic layer blocks the transmitted light ($T=0$) so that the absorbance $A$ is obtained only by the reflectance via $A=1-R$. However, this structure is not sensitive to the presence of a thin layer 1' on the MG because the electric field is weak and delocalized on the MG's interface between two slits (the input side). Another way to sense the minute variation of refractive index is by employing the perfect absorber made by replacing the MG with the metallic disk array supporting LSPs [24]. The figure of merit (FOM*) achieved by this structure reaches 87 1/RIU (RIU: refractive index unit) with water as a bulk sample. Here, we present the new perfect absorber with ultrahigh FOM* due to the tiny variation of the refractive index or thickness of a large-area thin film. Our optical structure is input/I/MG/M/output HP supporting the perfect absorption at the resonant frequency between the Wood's anomaly of the insulator and the Fabry-Perot (FP) of the slit. The designed resonant peak is extremely sharp with the minimum reflectance nearly zero resulting in the ultrahigh FOM*.

Figure 5(a) shows the schematic view of our structure where the input and output media are air, the insulator layer is the lossless HR-Si and the slit is filled with the lossy Zeonex polymer recently used as the host medium of the THz fibre hyperlens [27]. The HR-Si is chosen because it has low-frequency 1st-order Wood's anomaly due to its high dielectric constant, and Zeonex polymer is selected because it has low extinction coefficient thereby supporting ultra-sharp absorption peak. The absorbance of this structure for the normal incidence is obtained by using the following analytical reflection coefficient

$$\frac{B_m^{(\text{in})}}{H_0} = \exp\left(-ik_y^{(\text{in})}d_{1'}\right)\left[\frac{\exp\left(i\phi_0^{(1')}\right) + r_{\text{in},1'}^{(0)}\exp\left(-i\phi_0^{(1')}\right)}{\exp\left(-i\phi_0^{(1')}\right) + r_{\text{in},1'}^{(0)}\exp\left(i\phi_0^{(1')}\right)}\right]\delta_{m0}$$

$$+ \exp\left(-ik_y^{(\text{in})}d_{1'}\right)I_0 \frac{n_s \varepsilon_{1'}}{n_{1'}\varepsilon_s}iY_m^{(1')}$$

$$\times \frac{2t_{\text{in},1'}^{(0)}t_{1',\text{in}}^{(m)}}{\prod_{n=0,m}\left(\exp\left(-i\phi_n^{(1')}\right) + r_{\text{in},1'}^{(n)}\exp\left(i\phi_n^{(1')}\right)\right)}\langle \beta_m|0\rangle \frac{1}{G_{\infty s}^{(\text{in})} - \gamma}, \quad (12)$$

where the index 1' refers to the HR-Si layer. The dispersion relation is obtained by the condition $G_{\infty s}^{(\text{in})} - \gamma = 0$, where $G_{\infty s}^{(\text{in})}$ with $M=1$ is given by Eq. (13).

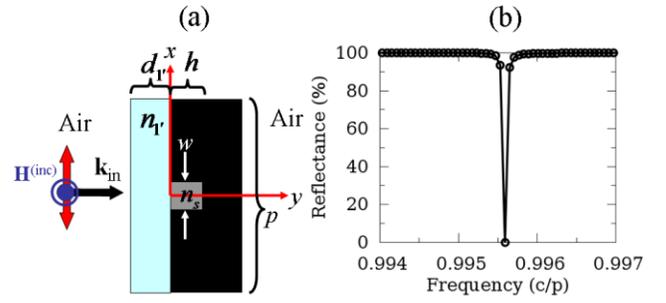

Fig. 5. (a) Schematic view of input/I/MG/M/output HP. The insulator layer 1' is lossless HR-Si with dielectric constant $\varepsilon_{1'}=11.7$, the slit is filled with lossy Zeonex polymer with dielectric constant $\varepsilon_s = 2.3104+0.002i$ [27], and the metallic layer 1 behind the MG is PEC. (b) Reflection spectrum of the input/I/MG/M/output HP giving the highest FOM* due to the presence of a sample layer with refractive index 1.01 and thickness $0.0001p$. The optimized HR-Si thickness and slit height are $0.018p$ and $0.934p$, respectively. The structure is excited by the normal incidence TM light.

$$G_{\infty s}^{(\text{in})} = i\frac{n_s \varepsilon_{1'}}{n_{1'}\varepsilon_s}\sum_{m=-\infty}^{\infty}\left(\frac{\exp\left(-i\phi_m^{(1')}\right) - r_{\text{in},1'}^{(m)}\exp\left(i\phi_m^{(1')}\right)}{\exp\left(-i\phi_m^{(1')}\right) + r_{\text{in},1'}^{(m)}\exp\left(i\phi_m^{(1')}\right)}\right)$$

$$\times Y_m^{(1')}\frac{w}{p}\text{sinc}^2(\beta_m w/2). \quad (13)$$

Note that if the layer 1' was air and $w\ll\lambda$, then $r_{\text{in},1'}^{(m)}=0$ and the dispersion relation would be reduced to that of the SSPPs on the metallic groove [28]. When a sample layer 2' is present next to the HR-Si layer, $\Delta n = n_{2'}-1 \neq 0$, $\Delta d = d_{2'} \neq 0$, and the reflection coefficient is changed to

$$\frac{B_m^{(\text{in})}}{H_0} = \left\{\frac{\exp\left(i\phi_0^{(+)}\right) + r_{\text{in},2'}^{(0)}\exp\left(-i\phi_0^{(+)}\right) + r_{2',1'}^{(0)}\left[\exp\left(i\phi_0^{(-)}\right) + r_{\text{in},2'}^{(0)}\exp\left(-i\phi_0^{(-)}\right)\right]}{\exp\left(-i\phi_0^{(+)}\right) + r_{\text{in},2'}^{(0)}\exp\left(i\phi_0^{(+)}\right) + r_{2',1'}^{(0)}\left[\exp\left(-i\phi_0^{(-)}\right) + r_{\text{in},2'}^{(0)}\exp\left(i\phi_0^{(-)}\right)\right]}\right\}\delta_{m0}$$

$$\times \exp\left[-ik_y^{(\text{in})}(d_{1'}+d_{2'})\right] + \exp\left[-ik_y^{(\text{in})}(d_{1'}+d_{2'})\right]I_0 \frac{n_s \varepsilon_{1'}}{n_{1'}\varepsilon_s}iY_m^{(1')}\langle\beta_m|0\rangle\frac{1}{G_{\infty s}^{(\text{in})}-\gamma}$$

$$\times \frac{2t_{\text{in},2'}^{(0)}t_{2',\text{in}}^{(m)}t_{2',1'}^{(0)}t_{1',2'}^{(m)}}{\prod_{n=0,m}\left[\exp\left(-i\phi_n^{(+)}\right)+r_{\text{in},2'}^{(n)}\exp\left(i\phi_n^{(+)}\right)+r_{2',1'}^{(n)}\left[\exp\left(-i\phi_n^{(-)}\right)+r_{\text{in},2'}^{(n)}\exp\left(i\phi_n^{(-)}\right)\right]\right]}$$

, (14)

where $\phi_n^{(+)} = \phi_n^{(2')} + \phi_n^{(1')}$, $\phi_n^{(-)} = \phi_n^{(2')} - \phi_n^{(1')}$ $(n=0,m)$, and the parameter $G_{\infty s}^{(\text{in})}$ is also changed to the new value given as follows

$$G_{\infty s}^{(\text{in})} = i\frac{n_s\varepsilon_{1'}}{n_{1'}\varepsilon_s}\sum_{m=-\infty}^{\infty}\left\{\frac{\exp\left(-i\phi_m^{(+)}\right) - r_{\text{in},2'}^{(m)}\exp\left(i\phi_m^{(+)}\right) - r_{2',1'}^{(m)}\left[\exp\left(-i\phi_m^{(-)}\right) - r_{\text{in},2'}^{(m)}\exp\left(i\phi_m^{(-)}\right)\right]}{\exp\left(-i\phi_m^{(+)}\right) + r_{\text{in},2'}^{(m)}\exp\left(i\phi_m^{(+)}\right) + r_{2',1'}^{(m)}\left[\exp\left(-i\phi_m^{(-)}\right) + r_{\text{in},2'}^{(m)}\exp\left(i\phi_m^{(-)}\right)\right]}\right\}$$

$$\times \frac{w}{p}\text{sinc}^2(\beta_m w/2). \quad (15)$$

Our design requires an unprecedented spectral resolution due to the rapid phase variation of the resonance near the Wood's anomaly. In order to optimize the FOM*, we define the spectral resolution as $\Delta f = 6\times 10^{-5}$ $c/p$, where the period $p$ is chosen as the length scale and the frequency is scalable with the unit $c/p$. For example, if $p$=300 μm, then $\Delta f = 60$ MHz. At the frequency $f \approx 1$ $c/p$, the spectral accuracy of our sensor is $\Delta f / f \approx 6\times 10^{-5}$. The backward-wave oscillator (BWO) generating the continuous THz wave has the spectral accuracy $\Delta f / f \approx 10^{-6}$ [29] which is sufficient to capture the resonant peak of our structure.

Figure 5(b) shows the 0th-order reflection spectrum of the optimized input/I/MG/M/output-HP having the highest FOM* which will be given in the next section. This structure has the HR-Si thickness $d_{1'} = 0.018 p$ and the slit height $h = 0.934 p$. The resonance occurs at the frequency 0.99559 $c/p$ which is slightly higher than the 1st-order Wood's anomaly of the HR-Si at 0.99426 $c/p$. The reflectance at the resonance is only 0.0003% ($A$=99.9997%). The spectral width of the resonant peak is smaller than the chosen spectral resolution. By defining the increased spectral resolution as $\Delta f = 1\times 10^{-6}$ $c/p$, the spectral width becomes $3.4\times 10^{-5}$ $c/p$.

Figure 5(b) shows the 0th-order reflection spectrum of the optimized input/I/MG/M/output-HP having the highest FOM* which will be given in the next section. This structure has the HR-Si thickness $d_{1'} = 0.018 p$ and the slit height $h = 0.934 p$. The resonance occurs at the frequency 0.99559 $c/p$ which is slightly higher than the 1st-order Wood's anomaly of the HR-Si at 0.99426 $c/p$. The reflectance at the resonance is only 0.0003% ($A$=99.9997%). The spectral width of the resonant peak is smaller than the chosen spectral resolution. By defining the increased spectral resolution as $\Delta f = 1\times 10^{-6}$ $c/p$, the spectral width becomes $3.4\times 10^{-5}$ $c/p$.

**B. Sensitivity of the input/I/MG/M-HP as the highly sensitive refractive index/thickness sensors of the large-area thin film**

The sensitivity of the perfect absorber sensor is generally characterized by the figure of merit (FOM*) defined in the practical way as

$$\text{FOM}^* = \left( \frac{(\Delta R / R)}{\Delta \eta} \right)_{\max}, \qquad (16)$$

where $R$ is the measurable reflectance of a sensor without a sample, $\Delta \eta$ denotes $\Delta n$ or $\Delta d$ of a sample layer, and $\Delta R$ is change of the reflectance. As $\Delta n \to 0$ or $\Delta d \to 0$, the FOM* takes the derivative form [24,25]. The asterisk differentiates the figure of merit obtained by monitoring the change of reflection from that obtained by monitoring the frequency shift [26]. We will adopt the FOM* because we believe that it truly harnesses the perfect absorption property.

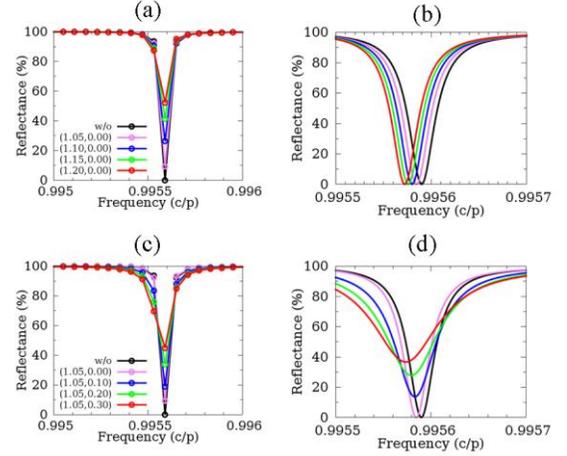

Fig. 6. (a) Reflection spectra of four sample layers deposited on the input/I/MG/M/output HP. The black points denote the sensor without (w/o) the sample layer. The refractive indices with neglected imaginary parts of the sample layers are indicated in the figure. (b) The zoom-in plot of Fig. 6(a) with the increased spectral resolution $\Delta f = 1 \times 10^{-6}$ $c/p$. (c) The reflection spectra of the sample layers with $\text{Re}(n_{2'}) = 1.05$ and $\text{Im}(n_{2'})$ =0.00, 0.10, 0.20, and 0.30 as indicated in the figure. (d) The zoom-in plot of Fig. 5(c) with the increased spectral resolution $\Delta f = 1 \times 10^{-6}$ $c/p$.

The FOM* of the optimized input/I/MG/M/output-HP in Fig. 5(b) due to the presence of the sample layer with refractive index $n_{2'} = 1.01$ and thickness $d_{2'} = 0.0001 p$ is equal to $1.49 \times 10^5$ 1/RIU. This means that the resonant reflectance increases $1.49 \times 10^3$ times from 0.0003% to 0.447% due to the presence of the large-area thin film on the perfect absorber.

The reflection spectra for several lossless sample layers with the same thickness $d_{2'} = 0.0001 p$ deposited on the optimized sensor (Fig. 5(b)) are shown in Fig. 6(a). The reflectance at the original resonance increases by increasing the real part of the refractive index. Indeed, the resonant position makes slightly red shift toward the lower frequency which can be seen in the zoom-in plot with the increased frequency resolution $\Delta f = 10^{-6}$ $c/p$ in Fig. 6(b) resulting in the increase of the reflectance at the original resonance. Notice that the spectral width of the original spectrum does not change and its reflection dip lies approximately at the same level because the sample is lossless. Figure 6(c) shows the reflection spectra of the lossy sample layers with $\text{Re}(n_{2'}) = 1.05$ and $\text{Im}(n_{2'})$ varied from 0.00 to 0.30. We also see in this figure that the reflectance at the original resonance increases by increasing $\text{Im}(n_{2'})$. Moreover, the spectra are broader and their associated reflection dips increase by increasing $\text{Im}(n_{2'})$ which can be clearly seen in the zoom-in plot with the increased spectral resolution $\Delta f = 10^{-6}$ c/p in Fig. 6(d).

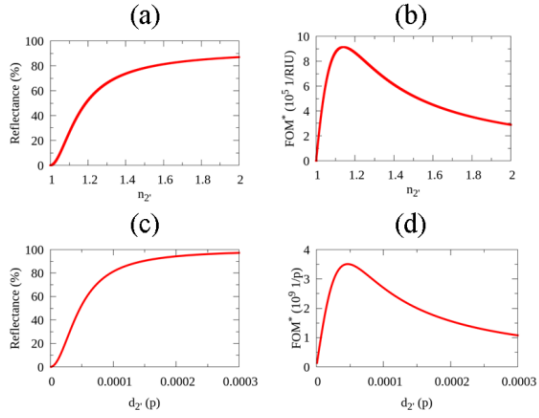

Fig. 7. (a) Reflectance, and (b) FOM* due to the presence of $\Delta n = n_{2'} - 1$ of a sample layer with thickness $d_{2'} = 0.0001p$ deposited on the optimized input/I/MG/M/output HP with $d_{1'} = 0.018p$ and $h = 0.934p$. (c) Reflectance, and (d) FOM* due to the presence of $\Delta d = d_{2'}$ of a sample layer with refractive index $n_{2'} = 1.6$ deposited on the optimized input/I/MG/M/output HP with $d_{1'} = 0.018p$ and $h = 0.934p$.

The reflectance is a nonlinear function of $n_{2'}$ and $d_{2'}$ of a sample layer which results in the non-constant FOM* relative to the sample's properties. Figure 7(a) shows the reflectance as a function of $n_{2'}$ with fixed $d_{2'} = 0.0001p$ of the sample layer deposited on top of the optimized sensor (Fig. 5(b)). By taking the slope of the reflection curve with respect to $n_{2'}$ in Fig. 7(a) and applying Eq. (16), we obtain the FOM* as a function of $n_{2'}$ shown in Fig. 7(b). The FOM* surges within the range $1 < n_{2'} < 1.14$ corresponding to the sharp increase of the slope of the reflection curve with respect to $n_{2'}$. Then, the FOM* reaches the maximum value about $9.13 \times 10^5$ 1/RIU at $n_{2'} = 1.14$ where it stays approximately constant over the narrow band. The maximum FOM* and the constant FOM* correspond to the maximum slope and the constant slope of the reflection curve, respectively. By increasing $n_{2'}$ further from the maximum point, the FOM* gradually decreases corresponding to the gradual decrease in the slope of the reflection curve. Next, we plot the reflectance as a function of $d_{2'}$ with fixed $n_{2'} = 1.6$ (a photoresist [26]) deposited on the optimized sensor (Fig. 5(b)) in Fig. 7(c), and the FOM* obtained by Fig. 7(c) with the application of Eq. (16) is shown in Fig. 7(d). The FOM* curve as a function of $d_{2'}$ has the same features and the same corresponding to the reflection curve as those due to the variation of $n_{2'}$, but the maximum value of FOM* is now about $3.5 \times 10^9$ 1/p which occurs at $d_{2'} = 4.5 \times 10^{-5} p$. This means that the reflectance increases $1.57 \times 10^5$ times from 0.0003% to 47% due to the presence of the photoresist with thickness $d_{2'} = 4.5 \times 10^{-5} p$. Therefore, this perfect absorber can be used to sense the presence of the large-are nano-film. For example, if p=300 μm, the photoresist with thickness 13.5 nm can be easily detected by using this perfect absorber sensor.

## C. Resonant properties, sensing mechanism, and future perspectives

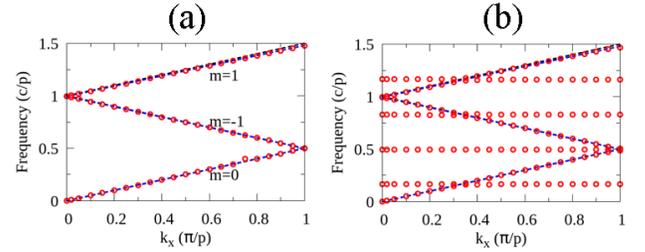

Fig. 8. Dispersion relation of the absorption resonances in the input/I/MG/M/output-HP with $d_{1'} = 0.018p$, $h = 0.02p$ (a), and $h = 0.934p$ (b). The black dashed lines denote the light lines in air (nearly overlap with the blue dashed line). The blue dashed lines denote the Wood's anomalies in HR-Si with their orders indicated in Fig. 8(a).

Next, we describe the physical origin of the resonance and the sensing mechanism of our sensor. Figure 8(a) and 8(b) show the dispersion relations of the absorption resonances in the $d_{1'} = 0.018p$ input/I/MG/M/output HP with the slit heights $h = 0.02p$ (un-optimized) and $h = 0.934p$ (optimized), respectively. The resonances of the $h=0.02p$ structure correspond to the SSPPs excited on the surface of HR-Si which neatly follow the mth-order Wood's anomaly in HR-Si determined by Eq. (11). The Wood's anomaly lines are close to the light lines in air because the HR-Si thickness is much smaller than the wavelength of light propagating in the insulator layer. For $k_x = 0$, the SSPP(HR-Si,m=1) is at 0.99355 c/p. By increasing the slit height, we expect that the cavity mode in the slit's cavity determined by the condition $\cos(n_s k_0 h) = 0$ will arise at the frequency $f_l^{(c)} = (2l-1)c/4n_s h$, and the Fabry-Perot (FP) determined by the condition $\sin(n_s k_0 h) = 0$ will arise at the frequency $f_l^{(FP)} = lc/2n_s h$, where l is an integer and "c" for cavity ("c" denotes the speed of light in vacuum). The difference between the cavity mode and the FP in this structure lies in the $\gamma$ parameter. The FP has $\gamma \to \infty$, and thus its diffraction term of the reflection coefficient (the second term in the right hand side of Eq. (12)) is vanishingly small which does not affect the non-diffraction term (the first term in the right hand side of Eq. (12)) thereby showing nearly perfect reflection. Therefore, the bands due to solely FP are not drawn in the dispersion relation diagram. On the other hand, the cavity mode has $\gamma \to 0$ reviving the diffraction term which leads to the interference between the non-diffraction and diffraction terms thereby generating the significant decrease of the reflection. For our optimized sensor with $h=0.934p$, we obtain the dispersion relation shown in Fig. 8(b). The flat bands correspond to the first four cavity modes which make anti-crossing with the SSPPs for non-zero $k_x$. We focus on the SSPP(HR-Si,$m$=1) at 0.99559 $c/p$ that gives the perfect absorption which attributes to the 1st-order Wood's anomaly at 0.99426 $c/p$ and the 3rd-order FP at 1.05654 $c/p$. The non-diffraction and diffraction terms at this frequency are approximately equal to $\exp(0.039\pi i)$ and $1.001\exp(-0.961\pi i)$, respectively. This means that the non-diffraction term is anti-parallel to the diffraction term with their amplitudes nearly the same. Therefore, the summation of these two terms leads to the extremely small reflectance about 0.0003% at this frequency. For the un-optimized $h=0.02p$ structure, the non-diffraction and diffraction terms of the SSPP(HR-Si,$m$=1) at 0.99355 $c/p$ are approximately equal to $\exp(0.039\pi i)$ and $0.402\exp(0.604\pi i)$, respectively. Therefore, the summation of these two terms leads to the high reflectance equal to 99.94% at this frequency. The main factor which results in the dramatic difference between the diffraction terms of two different slit heights is the dispersion function $\Omega = G_{\infty s}^{(in)} - \gamma$, according to Eq. (12). The ideal resonance is excited when the real and imaginary parts of $\Omega$ are simultaneously zero. However, this condition is hardly satisfied in reality. Therefore, what we calculate is the less stringent resonance condition: $\mathrm{Re}(\Omega) \approx 0$ and $\mathrm{Im}(\Omega) \approx 0$. The dispersion relations in Fig. 8(a)-(b) are computed by finding the set of pairs $(k_x, f)$ whose corresponding $\mathrm{Re}(\Omega)$ change sign by adding/subtracting frequency step and $|\mathrm{Re}(\Omega)|$ is smallest. The $\mathrm{Im}(\Omega)$ must be also small. Otherwise, the resonance will be forbidden. For the un-optimized structure in Fig. 8(a), the $\Omega$ of the SSPP(HR-Si,$m$=1) at $k_x$=0 and $f$=0.99355 $c/p$ is about $-7.4\times10^{-2}+1.5\times10^{-2}i$. For the optimized structure in Fig. 8(b), the $\Omega$ of the SSPP(HR-Si, $m$=1) at $k_x$=0 and $f$=99559 $c/p$ is about $-1.6\times10^{-5}+3.0\times10^{-2}i$. Therefore, the $\mathrm{Re}(\Omega)$ of the optimized structure is in three orders in magnitude smaller than that of the un-optimized structure. This leads to the dramatic difference in phase and magnitude of the diffraction terms from the optimized and un-optimized structure.

Next, we discuss the sensing mechanism of our sensor which can be understood by comparing the reflection coefficient of the sensor without the sample in Eq. (12) and that of the sensor with the sample in Eq. (14). It should be mentioned again that we measure only the 0th-order reflectance, thus $m$=0 is substituted into these two equations. When the large-area thin layer with thickness $d_{2'} = 0.0001p$ and refractive index $n_{2'} = 1.2$ is brought just next to the HR-Si layer of the optimized sensor, the non-diffraction and diffraction terms of the reflection coefficient in Eq. (12) are changed to those in Eq. (14). However, the non-diffraction term is slightly affected. The absolute value of the difference between the non-diffraction terms with and without the thin film is only 0.06% of the absolute value of the non-diffraction term without the thin film. Therefore, it is the diffraction term that is strongly affected by the presence of the thin film. In the diffraction term, the factor in front of $1/\Omega$ is approximately the same, while the $1/\Omega$ is dramatically altered. We find that the absolute value of $1/\Omega$ with the sample is about 28% of that without the sample. Therefore, the variation in $G_{\infty s}^{(in)}$ is responsible to the variation in the 0th-order reflectance ($\gamma$ is not varied). The $G_{\infty s}^{(in)}$ is large near the Wood's anomaly at which its denominator is closet to zero. This means that the variation of $G_{\infty s}^{(in)}$ at the resonance (which is close to the Wood's anomaly) results dominantly from the frequency shift of the Wood's anomaly because the denominator of the $G_{\infty s}^{(in)}$ is changed according to Eq. (13) and (15). In another word, the frequency shift of the Wood's anomaly is the main cause of the change of the reflection. The red shift of the Wood's anomaly due the presence of the sample is independent on the slit height. It is equal to about $2\times10^{-5}$ $c/p$ for our current sample deposited on the sensor with the HR-Si thickness $d_{1'} = 0.018p$.

The resonance is also shifted following the dislocation of the Wood's anomaly. Therefore, the high FOM* requires that (1) the frequency width of the resonant peak is comparable to or smaller than the frequency shift of the Wood's anomaly, and (2) the resonance is the perfect absorption. The latter is the most dominant factor. The resonant peak of our optimized sensor has the frequency width $3.4\times10^{-5}$ $c/p$ (obtained by the spectral resolution $10^{-6}$ $c/p$) which is comparable to the frequency shift of the Wood's anomaly, and the absorbance is 99.9997% at the resonant frequency 0.99559 $c/p$ which is nearly perfect absorption thereby giving the ultrahigh FOM*. For the un-optimized structure with $h=0.02p$, although the frequency width is about $6\times 10^{-6}$ $c/p$ (obtained by the spectral resolution $10^{-6}$ $c/p$) which is smaller than the frequency shift of the Wood's anomaly, but the absorbance is only 0.06% at the resonant frequency 0.99355 c/p. Therefore, the measured reflectance of the un-optimized structure is just changed to another nearly perfect reflection. This low absorption (high reflection) causes the extremely small FOM* about $6.2\times 10^{-4}$ 1/RIU due to the presence of the large-area thin film with thickness $d_{2'}=0.0001p$ and refractive index $n_{2'}=1.2$.

Our sensor can be improved in three main directions. The first direction is to reduce the effective wavelength of the SSPP(HR-Si,*m*=1) along the tangential axis . This means that the improved sensor will be able to sense the small-area thin layer such as bacteria or viruses. This can be done by working in lower frequency with the SSPP(HR-Si,*m*=0) lying outside the light cone which then allows us to choose smaller grating period. The SSPP(HR-Si,*m*=0) can be excited by using two MGs: one for in-coupling and another for supporting the SSPP(HR-Si,*m*=0). The analytical solutions reported here cannot be readily applied to this structure, but they may be useful to develop this more advanced structure. The second direction is to generate the electrical readout. A good sensor should be able to convert the variation in Δ*n* or Δ*d* interrogated by THz light into the electrical current. This may be achieved by combining graphene to the sensor and use the photothermoelectric effect in graphene [31]. The third direction is to make the sensor tunable by the application of an external stimuli for sensing over broad frequency range. This may be done by integrating the sensor with graphene surface plasmons [32], graphene metamaterial [33], and magnetic materials [34]. Nonetheless, our input/I/MG/M/output-HP shows that the presence of a smooth and large-area nano-film can be detected by properly choosing its parameters, an important step toward the practical THz thin-film sensing. Another promising structure that can be used to sense the nano-film with THz radiation is the annular gap array [35] which detect the frequency shift rather than the intensity variation due to the variation of local refractive index. This structure does not require high spectral resolution and thus the terahertz time-domain spectroscopy (THz-TDS) is sufficient for the characterization. However, unlike our sensor this refractive index sensor is not truly flat, and the sample is assumed to be inside the annular gap to perturb the local refractive index thereby shifting the resonant frequency. It is also needed to be improved in the same directions as our proposed sensor, and its analytical solutions are still missing.

## 4. Conclusion

In summary, we have given the QANS of the HP comprising the MG with narrow slit width and SDLs attached on two sides of the MG for THz radiation. The QANS agree with the FDTD simulation when the slit width is much smaller than the light wavelength so that the effects of the LSPs at the slit corners are negligible. It is shown that the high-order SSPPs excited on the surface of the dielectric layers with strong electric field enhancement are induced by the Wood's anomalies. The QANS are applied to optimize the perfect absorber whose 1st-order Wood's anomaly of the insulator layer and the FP in the slit's cavity account for the perfect absorption for refractive index and thickness sensing of large-area thin film. The figures of merit depend on properties of the large-area thin film and reach fifth order and ninth order in magnitude due to the variations of refractive index and thickness, respectively, of the thin film. The QANS and the new perfect absorber give the tools for the optimization of the practical thin-film refractive index sensor for THz radiation.

## References

1. M. Tonouchi, "Cutting-edge terahertz technology," Nat. Photon. **1**, 97-105 (2007).
2. S. Jae Oh, S.-H. Kim, Y. B. Ji, K. Jeong, Y. Park, J. Yang, D. W. Park, S. K. Noh, S.-G. Kang, Y.-M. Huh, J.-H. Son, and J.-S. Sue, "Study of freshly excised brain tissues using terahertz imaging," Biomed. Opt. Express **5**(8), 2837-2842 (2014).
2. B. Zeng, Y. Gao, and F. J. Bartoli, "Rapid and highly sensitive detection using Fano resonances in ultrathin plasmonic nanogratings," Appl. Phys. Lett. **105**, 161106 (2014).
4. F. Miyamaru, S. Hayashi, C. Otani, and K. Kawase, "Terahertz surface-wave resonant sensor with a metal hole array," Opt. Lett. **31**(8), 1118-1120 (2006).
5. F. D'Apuzzo, P. Candeloro, F. Domenici, M. Autore, P. Di Pietro, A. Perucchi, P. Roy, S. Sennato, F. Bordi, E. M. Di Fabrizio, S. Lupi, "Resonating terahertz response of periodic arrays of subwavelength apertures," Plasmonics **10**, 45-50 (2015).
6. T. W. Ebbesen, H. J. Lezec, H. F. Ghaemi, T. Thio, and P. A. Wolf, "Extraordinary optical transmission through sub-wavelength hole array," Nature **391**, 667-669 (1998).
7. J. A. Porto, F. T. Garcia-Vidal, and J. B. Pendry, "Transmission resonances on metallic gratings with very narrow slits," Phys. Rev. Lett. **83**, 2845-2848 (1999).
8. J. B. Pendry, L. Martin-Moreno, and F. J. Garcia-Vidal, "Mimicking surface plasmons with structured surfaces," Science **305**, 847-848 (2004).
9. A. E. Cetin, D. Etezadi, B. C. Galarreta, M. P. Busson, Y. Eksioglu, and H. Altug, "Plasmonic nanohole arrays on a robust hybrid substrate for highly sensitive label-free biosensing," ACS Photonics **2**, 1167-1174 (2015).
10. B. Caballero, A. Garcia-Martin, and J. C. Cuevas, "Hybrid magnetoplasmonic crystals boost the performance of nanohole arrays as plasmonic sensors," ACS Photonics **3**, 203-208 (2016).
11. F. J. Garcia-Vidal, L. Martin-Moreno, T. W. Ebbesen, and L. Kuiper, "Light passing through subwavelength apertures," Rev. Mod. Phys. **82**(1), 729-787 (2010).
12. S. Ishii, A. V. Kildishev, E. Narimanov, V. M. Shalaev, and V. P. Drachev, "Sub-wavelength interference pattern from volume plasmon polaritons in a hyperbolic medium," Laser & Photon Rev. **7**(2), 265-271 (2013).

13. A. F. Oskooi, D. Roundy, M. Ibanescu, P. Bermel, J. D.Joannopoulos, and S. G. Johnson, "Meep: A flexible free-software package for electromagnetic simulations by the FDTD method," Comput. Phys. Commun. **181**, 687-702 (2010).
14. M. A. Ordal, L. L. Long, R. J. Bell, S. E. Bell, R. R. Bell, R. W. Alexander, Jr., and C. A. Ward, "Optical properties of the metals Al, Co, Cu, Au, Fe, Pb, Ni, Pd, Pt, Ag, Ti, and W in the infrared and far infrared," Appl. Opt. **22**(7), 1099-1120 (1983).
15. D Grischkowsky, S. Keiding, M. V. Exter, and C. Fattinger, "Far-infrared time-domain spectroscopy with terahertz beams of dielectrics and semiconductors," J. Opt. Soc. Am. B **7**(10), 2006-2015 (1990).
16. Y. S. Jin, G. J. Kim, and S. G. Jeon, "Terahertz dielectric properties of polymers," JKPS **49**(2), 513-517 (2006).
17. A. Hessel and A. A. Oliner, "A new theory of Wood's anomalies on optical gratings," Appl. Opt. **4**(10), 1275-1297 (1965).
18. Lord Rayleigh, "On the dynamical theory of gratings," Proc. Roy. Soc. (London) **A79**, 399-416 (1907).
19. G. D. Aguanno, N. Mattiucci, A. Alu, and M. J. Bloemer, "Quenched optical transmission in ultrathin subwavelength plasmonic gratings," Phys. Rev. B **83**, 035426 (2011).
20. K. A. Willets and R. P. V. Duyne, "Localized surface plasmon resonance spectroscopy and sensing," Annu. Rev. Phys. Chem. **58**, 267-297 (2007).
21. D. M. Pozar, Microwave Engineering, Wiley, USA, (2012).
22. M. A. Seo, H. R. Park, S. M. Koo, D. J. Park, J. H. Kang, O. K. Suwal, S. S. Choi, P. C. M. Planken, G. S. Park, N. K. Park, Q. H. Park and D. S. Kim, "Terahertz field enhancement by a metallic nano slit operating beyond the skin-depth limit," Nat. Photon. **3**, 152-156 (2009).
23. A. P. Hibbins and J. R. Sambles, "Squeezing millimeter waves into microns," Phys. Rev. Lett. **92**, 143904 (2004).
24. N. Liu, M. Mesch, T. Weiss, M. Hentschel, and H. Giessen, "Infrared perfect absorber and its application as plasmonic sensor," Nano Lett. **10**, 2342-2348 (2010).
25. J. Becker, A. Trügler, A. Jakab, U. Hohenester, and C. Sönnichsen, "The optimal aspect ratio of gold nanorods for plasmonic bio-sensing," Plasmonics 5, 161-167 (2010).
26. L. Cong, S. Tan, R. Yahiaoui, F. Yan, W. Zhang, and R. Singh, "Experimental demonstration of ultrasensitive sensing with terahertz metamaterial absorbers: a comparison with the metasurfaces," Appl. Phys. Lett. **106**, 031107 (2015).
27. A. Tuniz, K. J. Kaltenecker, B. M. Fischer, M. Walther, S. C. Fleming, A. Argyros, and B. T. Kuhlmey, "Metamaterial fibres for subdiffraction imaging and focusing at terahertz frequencies over optically long distances," at. Commun. **4**, 2706 (2013).
28. F. J. Garcia-Vidal, L. Martin-Moreno, and J. B. Pendry, "Surfaces with holes in them: new plasmonic metamaterials," . Opt. A:Pure Appl. Opt. **7**, S97-S101 (2005).
29. B. Gompf, M. Gerull, T. Muller, and M. Dressel, "THz-micro-spectroscopy with backward-wave oscillators," Inf. Phys. & Tech. 49, 128-132 (2006).
30. P. Törmä, and W. L. Barnes, "Strong coupling between surface plasmon polaritons and emitters: a review," Rep. Prog. Phys. **78**, 013901 (2015).
31. X. Cai, A. B. Sushkov, R. J. Suess, M. M. Jadidi, G. S. Jenkins, L. O. Nyakiti, R. L. Myers-Ward, S. Li, J. Yan, D. K. Gaskill, T. E. Murphy, H. D. Drew, and M. S. Fuhrer, "Sensitive room-temperature terahertz detection via the photothermoelectric effect in graphene," Nat. Nano. **9**, 814-819 (2014).
32. M. M. Jadidi, A. B. Sushkov, R. L. Myers-Ward, A. K. Boyd, K. M. Daniels, D. K. Gaskill, M. S. Fuhrer, H. D. Drew, and T. E. Murphy, "Tunable terahertz hybrid metal-graphene plasmons," Nano Lett. **15**, 7099-7104 (2015).
33. L. Ju, B. Geng, J. Horng, C. Girit, M. Martin, Z. Hao, H. A. Bechtel, X. Liang, A. Zettl, Y. R. Shen, and Feng Wang, "Graphene plasmonics for tunable terahertz metamaterials," Nat. Nano. **6**, 630-634 (2011).
34. V. V. Temnov, G. Armelles, U. Woggon, D. Guzatov, A. Cebollada, A. Garcia-Martin, J.-M. Garcia-Martin, T. homay, A. Leitenstorfer, and R. Bratschitsch, "Active magneto-plasmonics in hybrid metal–ferromagnet structures," Nat. Photon. **4**, 107-111 (2010).
35. H.-R. Park, X. Chen, N.-C. Nguyen, J. Peraire, and S.-H. Oh, "Nanogap-enhanced terahertz sensing of 1 nm thick (λ/106) dielectric films," ACS Photon. **2**, 417-424 (2015).